\documentclass{kapproc} 
\usepackage{procps} 
\usepackage[dvips]{graphicx}
\upperandlowercase
\kluwerbib 

\def\A&A{\em A\&A}

\def\Msun{M$_{\odot}$}

\begin{document}

\articletitle{The Mass Spectrum of X-Ray Binaries}


\author{Jorge Casares}
\affil{Instituto de Astrofisica de Canarias. 38200- La Laguna. 
Tenerife. Spain}
\email{jcv@iac.es}

\begin{abstract}
This review summarizes the observational constraints on the mass 
spectrum of compact objects in X-ray binaries. We currently have 20 X-ray 
binaries with confirmed black holes, based on dynamical information (i.e. mass 
in excess of 3 \Msun). In two cases, V404 Cyg and GRS 1915+105, the black
hole mass exceeds the maximum predicted by current Type Ib 
supernovae models and challenges black hole formation scenarios. 
The great majority of black hole binaries are members of the class of X-ray 
Transients, where long periods of quiescence enable spectroscopic studies of 
the faint donor stars. On the other hand, neutron star binaries are mostly 
found in persistent binaries, where reprocessed light from the accretion disc 
overwhelms the companion star and precludes mass estimates. 
New results, based on the detection of optical fluorescent lines from 
the donor star and X-ray burst oscillations, provide the best prospects 
for mass constraints of neutron stars in persistent X-ray binaries.     
\end{abstract}


\section*{Introduction}
Building the mass distribution of compact objects is a fundamental
experiment in modern Astrophysics which can only be done in X-ray
binaries. The interest of this research is two-fold: set constraints
on the equation of state of nuclear matter (hereafter EOS) and test
models of supernovae explosions and close binary evolution.  X-ray
binaries are interacting binaries where a normal star transfers matter
onto a compact object, a black hole (BH) or neutron star (NS).  Matter
is accelerated in the strong gravitational field of the compact star
and heated up to $\sim 10^7$ K before being accreted. This is the
canonical model, first proposed by ~\cite{shk67} to explain the new
X-ray sources detected in the 60's and 70's by X-ray satellites such as 
UHURU or Einstein. In the 80's and 90's, a new generation of higher
sensitivity satellites, such as Ginga, Rosat and XTE, revealed a
population of several hundred X-ray binaries in the Galaxy with X-ray
luminosities in the range 10$^{36}$ -- 10$^{39}$ ergs s$^{-1}$ and
displaying rapid variability, down to the kilohertz regime. And
nowadays, the large collecting area and high angular resolution of
satellites such as Chandra and XMM-Newton allow us to resolve the X-ray
binary population in nearby galaxies, obtaining the X-ray luminosity
function and proving that they are responsible for typically a third
of the galaxy's total X-ray emission.

X-ray binaries are classically divided into two populations, based on the nature 
of the optical (companion) stars: high mass X-ray binaries (HMXBs) 
and low mass X-ray binaries (LMXBs). In HMXBs the companion is a 
hot luminous O-B supergiant whose bolometric luminosity completely dominates 
the energy spectrum. These binaries are relatively young (short 
lived) with estimated ages in the range $10^7-10^8$ yrs. There are about 
$\sim$ 100 such binaries in the 
Galaxy, strongly concentrated towards the spiral arms and, hence, they are 
considered as good tracers of the star formation rate  
(\cite{grimm02}). Neutron stars in HMXBs are recognized by the presence of 
regular X-ray pulsations as a consequence of their strong magnetic fields. 

On the other hand, LMXBs contain low-mass companion stars of spectral
types (mainly) K-M. These binaries are very compact, with orbital
periods clustering around 4-8 hours and concentrated in location
towards the galactic bulge. They are associated with the old stellar
population and show a large spread in galactic latitude, interpreted
as a signature of kick velocities received when the supernova
explosion formed the compact star. Neutron stars in LMXBs are also
revealed by the exhibition of sporadic X-ray bursts i.e.
thermonuclear eruptions of matter slowly accreted over the surface.
There are $\sim$ 200 X-ray active or "persistent" sources in the
Galaxy, and about $10^3$ "transients", which only show X-ray activity
occasionally.  The reason for this transient behaviour is the mass
transfer rate from the donor star $\dot{M_2}$, which is driven by the
binary/donor evolution. In X-ray transients, $\dot{M_2}$ is lower than
a critical value $\sim$ 10$^{-9}$ \Msun yr$^{-1}$ and this triggers
thermal-viscous instabilities in the accretion disc. This causes
enhanced mass transfer episodes onto the compact object, the so-called
"outbursts", with recurrence times of a few decades (\cite{king99}).
In the interim, these systems remain in the "quiescent" state, with
typical X-ray luminosities below $\sim 10^{32}$ ergs s$^{-1}$. It is during
these periods when we can attempt to detect the faint low-mass star
and extract dynamical information. The many studies performed on
quiescent X-ray transients have demonstrated that BHs outnumber NS by
more than 70\%, and hence the transients provide excellent hunting
grounds for BHs. More details about the optical and X-ray properties
of galactic X-ray binaries can be found in \cite{char04}, \cite{mcc04}
and \cite{psa04}.

\section{Establishing Black Holes}

As opposed to NS, which often show X-ray bursts or pulses, the best 
observational evidence for a BH is still its mass function $f(M_{\rm x})$. 
This equation relates the masses of the compact object $M_{\rm x}$, the companion 
star $M_{\rm c}$ (or alternatively the binary mass ratio 
$q=M_{\rm c}/M_{\rm x}$) and the binary inclination angle $i$ with two 
quantities to be extracted directly from the radial velocity curve of the 
companion star: the orbital period $P_{\rm orb}$ and 
the radial velocity semi-amplitude $K$: 

\begin{equation}
f(M) =  {{K^{3} P_{\rm orb}}\over 2 \pi G} = 
{{M^{3}_{\rm x}\sin^{3} i}\over \left(M_{\rm x} + M_{\rm c}\right)^2} = 
{{M_{\rm x} \sin^3 i}\over \left(1+q\right)^2}
\label{eq1}
\end{equation}

Since $M_{\rm c}>0$ and $0<i<90^{\circ}$ it is straightforward to show that 
$f(M)$ is a lower limit on $M_{\rm x}$. Therefore, a 
mass function larger than 3 \Msun, i.e. the maximum gravitational mass of 
a NS (\cite{rho74}), is taken as a secure proof for the existence of 
a BH, independently of the actual values of $i$ and $q$.

Table 1 presents an updated list of confirmed BHs based on this simple dynamical 
argument. We currently have 20 BHs, with orbital periods ranging from 4.1 hours 
to 33.5 days. 
There are 17 transient LMXBs and only 3 persistent systems, the 
HMXBs Cyg X-1 and two sources from the Large Magellanic Clouds: LMC X-1 and 
LMC  X-3. Note that, in addition to the three HMXBs, six transient LMXBs  
have mass functions < 3 \Msun. However, we have solid constraints on the 
inclination and the companion's mass for these binaries which result in 
M$_{\rm x}$ > 3 \Msun. 
The last one added to the list, BW Cir, contains an evolved G5 donor star in 
a 2.5 day orbit. Its optical luminosity 
places the binary at a distance $\ge$ 27 kpc and makes BW Cir the furthest BH 
binary in the Galaxy yet. Remarkably, 
its large systemic velocity (103 km s$^{-1}$) is in good agreement with the 
projected velocity of the Galactic differential rotation at that distance 
(see more details in \cite{casa04a}).

\begin{table}[ht]
\caption[Confirmed black holes and mass determinations]
{Confirmed black holes and mass determinations}
\begin{tabular*}{\textwidth}{@{\extracolsep{\fill}}lccccc}
\sphline
\it System & \it P$_{orb}$ & \it  f(M) & \it Donor  & \it Classification &
\it M$_{\rm x}$ \cr
&  \it (days) & \it (\Msun) & \it Spect. Type & & \it (\Msun) \cr
\sphline
GRS 1915+105    &     33.5    &    9.5 $\pm$ 3.0        &    K/M III   &  LMXB/Transient  & 14 $\pm$ 4  \cr
V404 Cyg        &      6.470  &   6.08 $\pm$ 0.06       &    K0 IV     &      ,,      & 12 $\pm$ 2 \cr
Cyg X-1         &      5.600  &   0.244 $\pm$ 0.005     &    09.7 Iab  &  HMXB/Persistent & 10 $\pm$ 3 \cr
LMC X-1         &      4.229  &   0.14 $\pm$ 0.05       &    07 III    &      ,,      & > 4 \cr
XTE J1819-254   &      2.816  &   3.13 $\pm$ 0.13       &    B9 III    &  LMXB/Transient  & 7.1 $\pm$ 0.3 \cr 
GRO J1655-40    &      2.620  &   2.73 $\pm$ 0.09       &    F3/5 IV   &      ,,      & 6.3 $\pm$ 0.3 \cr
BW Cir~$^a$      &      2.545  &   5.75 $\pm$ 0.30       &    G5 IV     &      ,,      & $>$ 7.8 \cr	 
GX 339-4        &      1.754  &   5.8  $\pm$ 0.5        &     --       &      ,,      &   \cr
LMC X-3         &      1.704  &   2.3  $\pm$ 0.3        &    B3 V      &  HMXB/Persistent & 7.6 $\pm$ 1.3\cr
XTE J1550-564   &      1.542  &   6.86 $\pm$ 0.71       &    G8/K8 IV  &  LMXB/Transient  & 9.6 $\pm$ 1.2 \cr
4U 1543-475     &      1.125  &   0.25 $\pm$ 0.01       &    A2 V      &      ,,      & 9.4 $\pm$ 1.0 \cr
H1705-250       &      0.520  &   4.86 $\pm$ 0.13       &    K3/7 V    &      ,,      & 6 $\pm$ 2 \cr
GS 1124-684     &      0.433  &   3.01 $\pm$ 0.15       &    K3/5 V    &      ,,      & 7.0 $\pm$ 0.6 \cr
XTE J1859+226~$^b$   &  0.382  &   7.4  $\pm$ 1.1        &     --       &      ,,      &     \cr
GS2000+250      &      0.345  &   5.01 $\pm$ 0.12       &    K3/7 V    &      ,,      & 7.5 $\pm$ 0.3 \cr
A0620-003       &      0.325  &   2.72 $\pm$ 0.06       &    K4 V      &      ,,      & 11 $\pm$ 2 \cr
XTE J1650-500   &      0.321  &   2.73 $\pm$ 0.56       &    K4 V      &      ,,      &   \cr
GRS 1009-45     &      0.283  &   3.17 $\pm$ 0.12       &    K7/M0 V   &      ,,      & 5.2 $\pm$ 0.6 \cr
GRO J0422+32    &      0.212  &   1.19 $\pm$ 0.02       &    M2 V      &      ,,      & 4 $\pm$ 1   \cr
XTE J1118+480~$^c$   &  0.171  &   6.3  $\pm$ 0.2        &    K5/M0 V   &      ,,      &  6.8 $\pm$ 0.4 \cr
\sphline
\end{tabular*}
\begin{tablenotes}
$^a$ The 1-year alias period at 2.564 days is equally significant. 
In this case the BH would be even 
strengthen with $f(M_{\rm x})=6.60 \pm 0.36$ \Msun ~(see \cite{casa04a}). \\
$^b$ Period is uncertain, with another possibility at 0.319 days (see
\cite{zur02}). This would drop the mass function to $f(M_{\rm x})=6.18$ \Msun.  \\
$^c$ Mass function updated after \cite{torres04}.
\end{tablenotes}
\end{table}
\inxx{captions,table}

\noindent

In addition to the mass function, one obviously needs a knowledge of the 
inclination and the mass ratio to fully determine the masses of the two stars.   
In the case of a BH binary, we face a "single-line spectroscopic binary" 
problem and all the information must be extracted from the optical companion.
However, there is a complete solution to the problem which involves:  
(i) the determination of the binary mass ratio through the rotational  
broadening $V_{\rm rot} \sin~ i$ of the companion's absorption lines and (ii) 
the determination of the inclination angle by fitting synthetic models to 
ellipsoidal lightcurves. This is the classic method to derive masses.  
Further details of these techniques and the systematics involved can be found 
in several review papers e.g. \cite{casa01}, \cite{char04}.  

\begin{figure}[ht]
\begin{center}
\begin{picture}(250,230)(50,30)
\put(0,0){\includegraphics{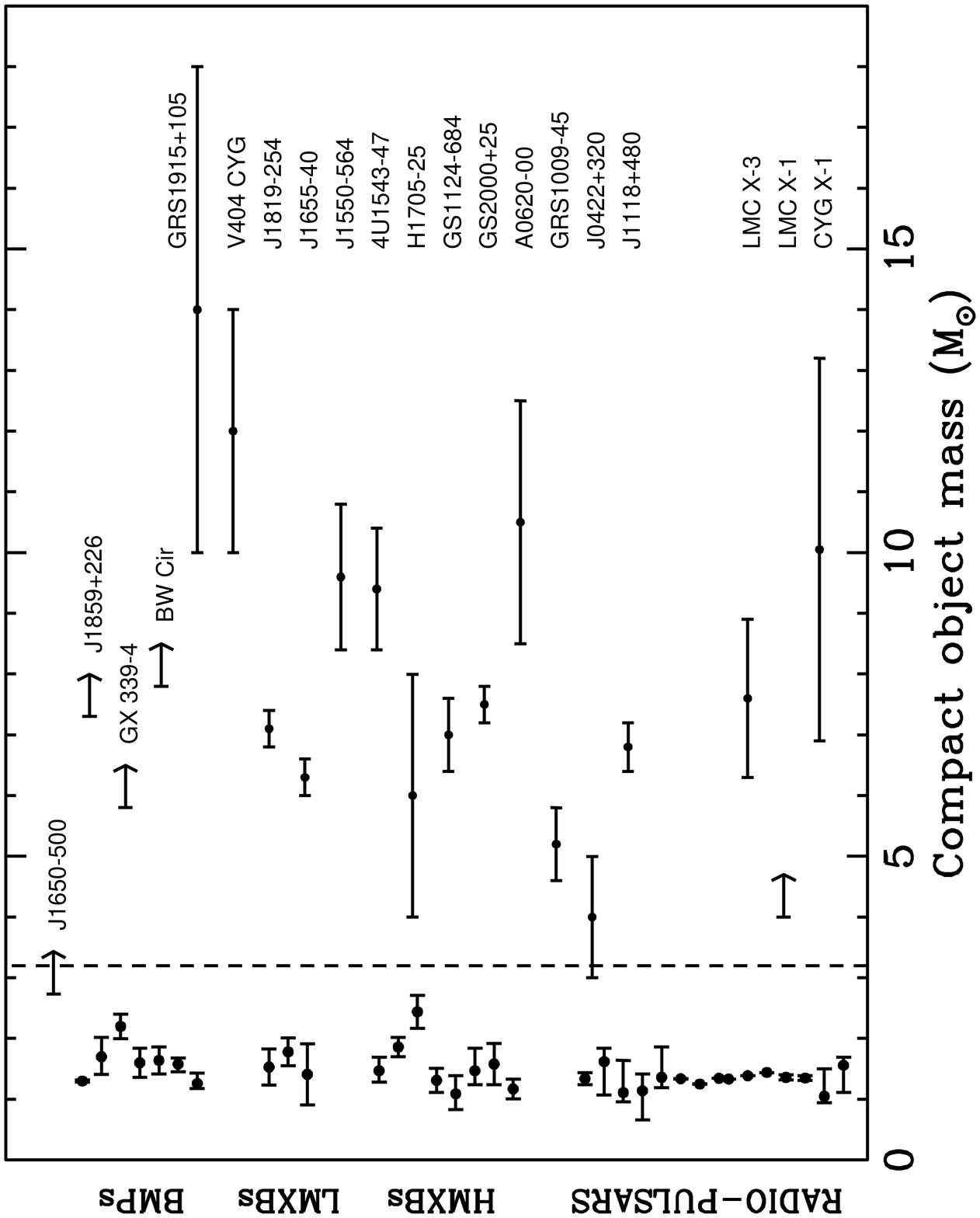}}
\noindent
\end{picture}
\end{center}
\caption{Mass distribution of compact objects in X-ray binaries. The arrows
indicate lower limits to BH masses. The dotted vertical line marks the maximum
mass allowed for a stable NS.}
\label{fig1}
\end{figure}

Following this prescription, we currently have 15 reliable BH masses which 
are listed in the last column of Table 1, updated from the compilations in 
\cite{oro03} and \cite{char04}. Figure 
1 plots these masses with their 1$\sigma$ errorbars. BH masses spread 
between 4 and 14 \Msun, with typical uncertainties in the range 5-33\%. 
Also in Fig. 1 we show 33 well determined NS masses, extracted from 
\cite{stairs04} and \cite{latt04}, also with 1$\sigma$ uncertainties.   
The most precise NS masses have been measured in a group of binary 
radio-pulsars. They are descendants of HMXBs, composed of two young pulsars 
whose orbits are known to great accuracy from pulse time delays. Relativistic  
effects lead to NS mass determinations with exquisite accuracy and they display 
a normal distribution centered at the canonical value of 
1.35 \Msun ~with a very small dispersion of $\pm 0.04$ \Msun. 
Dynamical masses are also available from pulsing NS in seven HMXBs, 
six of which are eclipsing. 
However, the uncertainties are much larger because the 
radial velocities of the optical stars are distorted by non-Keplerian
perturbations, caused by their strong winds. A few measurements also 
exist for LMXBs and binary millisecond pulsars (BMPs). BMPs are descendants of 
LMXBs, composed of a millisecond NS (spun up by accretion) and a detached 
white dwarf. Both LMXBs 
and BMPs are potential sites for massive NS because binary evolution predicts 
accretion of several tenths of a solar mass during their lifetimes. 

A fundamental result for the understanding of nuclear matter would be to 
find a NS more massive than 1.6 \Msun~ 
since this would rule out soft EOS (\cite{brown94}). Currently the best 
candidates are found in the LMXB Cyg X-2 (1.78 $\pm$ 0.23 \Msun:
\cite{casa98}, \cite{oro99}), the BMP J0751+1807 (2.2 $\pm$ 0.2
\Msun: \cite{nice04}) and the HMXBs 4U 1700-37 (2.44 $\pm$ 0.27 
\Msun: \cite{clark02}) and Vela X-1 (1.86 $\pm$ 0.16 \Msun:
\cite{barziv01}). 
The latter case, however, is less secure since the radial velocity curve of 
the companion is affected by systematic excursions which prevents a 
confirmation of the mass estimate (see ~\cite{barziv01}). 
In the case of J0751+1807 and 4U 1700-37, soft EOS are ruled out even at the 
95\% level. 

Very recently, large NS masses have also been reported for 2S0921-630 (=V395
Car) by two different groups: 2.0--4.3 \Msun ~(\cite{sha04}) and 
1.9--2.9 \Msun ~(\cite{jonker04}). However, the compact object in this LMXB 
has never shown any evidence for X-ray bursts or pulsations, 
so it could be a low-mass BH. This is also the situation for the 
compact object in 4U 1700-37, since it has never shown any NS signature. 

\begin{figure}[ht]
\begin{center}
\begin{picture}(250,200)(50,30)
\put(0,0){\includegraphics{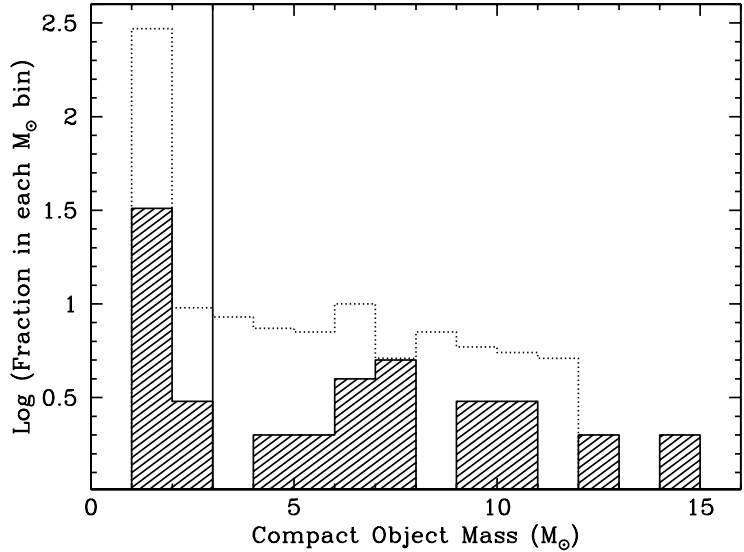}}
\noindent
\end{picture}
\end{center}
\caption{Observed mass distribution of compact objects in X-ray binaries 
(shaded histogram), compared to the theoretical distribution computed in 
Fryer \& Kalogera (2001) for "Case C + Winds" scenario (dotted line). The 
model distribution has been re-scaled for clarity.}
\label{fig2}
\end{figure}

Figure 2 presents the histogram of compact object masses compared to the 
theoretical distribution of remnants computed in \cite{fryer01} for the 
case of binary interaction under Case C mass transfer (i.e. Common Envelope
evolution after core helium ignition) and mass loss through winds in the 
Wolf-Rayet phase. This is the most realistic scenario since evolution through
Case B mass transfer (i.e. Common Envelope and H envelope removal before core 
helium ignition) fails to
produce BH remnants $>$ 3 \Msun~ with the current Wolf-Rayet mass loss rates 
(\cite{woos95}). The model predicts a BH mass cut at 12 \Msun~ which is 
difficult to reconcile with the high masses measured in V404 Cyg (12 
$\pm$ 2 \Msun) and GRS 1915+105 (14 $\pm$ 4 \Msun). Despite the large 
uncertainties in the masses, these two X-ray binaries seem to pose a challenge 
to BH formation theories and, in particular, suggest that mass loss rates in 
the WR phase are overestimated. 

Our histogram also shows a shortage of objects at 3--4 \Msun, not
predicted by the model distribution. If the gap is eventually
confirmed, it could strongly restrict the supernova explosion energy
since, as explained in ~\cite{fryer01}, it can be reproduced by a step
function dependence with the progenitor's mass. However, selection
effects could also be playing a role here since low mass BHs are
likely to show up as persistent X-ray sources, where dynamical masses
are difficult to obtain. The case of 2S0921-630 could well be an
example.  Obviously we are limited by low number statistics in the
observed distribution of compact remnants. Clearly more X-ray
transient discoveries and lower uncertainties in the mass
determinations are required before these issues can be addressed and the
form of the distribution can be used to constrain supernova models and X-ray
binary evolution.

\section{Mass Determination in Persistent LMXBs}

So far we have been dealing with mass determination in quiescent X-ray binaries. 
Dynamical studies in persistent LMXBs are, on the other hand, hampered by 
the huge optical luminosity of the accretion disc. This is driven by 
reprocessing of the powerful (Eddington limited) X-ray luminosity and 
completely swamps the spectroscopic features of the faint companion stars. 

New prospects for mass determination have been opened by the discovery of
high-excitation 
emission lines arising from the donor star in Sco X-1 (~\cite{stee02}). 
The most prominent are found in the core of the Bowen blend, namely the 
triplets NIII $\lambda$4634-40 and CIII $\lambda$4647-50.  
In particular, the NIII lines are powered by fluorescence resonance which 
requires seed photons of HeII Ly-$\alpha$. 
These narrow components move in phase with each other and are not resolved 
(FWHM=50 km s$^{-1}$ i.e. the instrumental 
resolution), an indication that the reprocessing region is very localized 
(Fig. 3). The extreme narrowness rules out the accretion flow or the hot 
spot and points to the companion star as the reprocessing site.   

\begin{figure}[ht]
\begin{center}
\begin{picture}(250,130)(50,30)
\put(0,0){\includegraphics{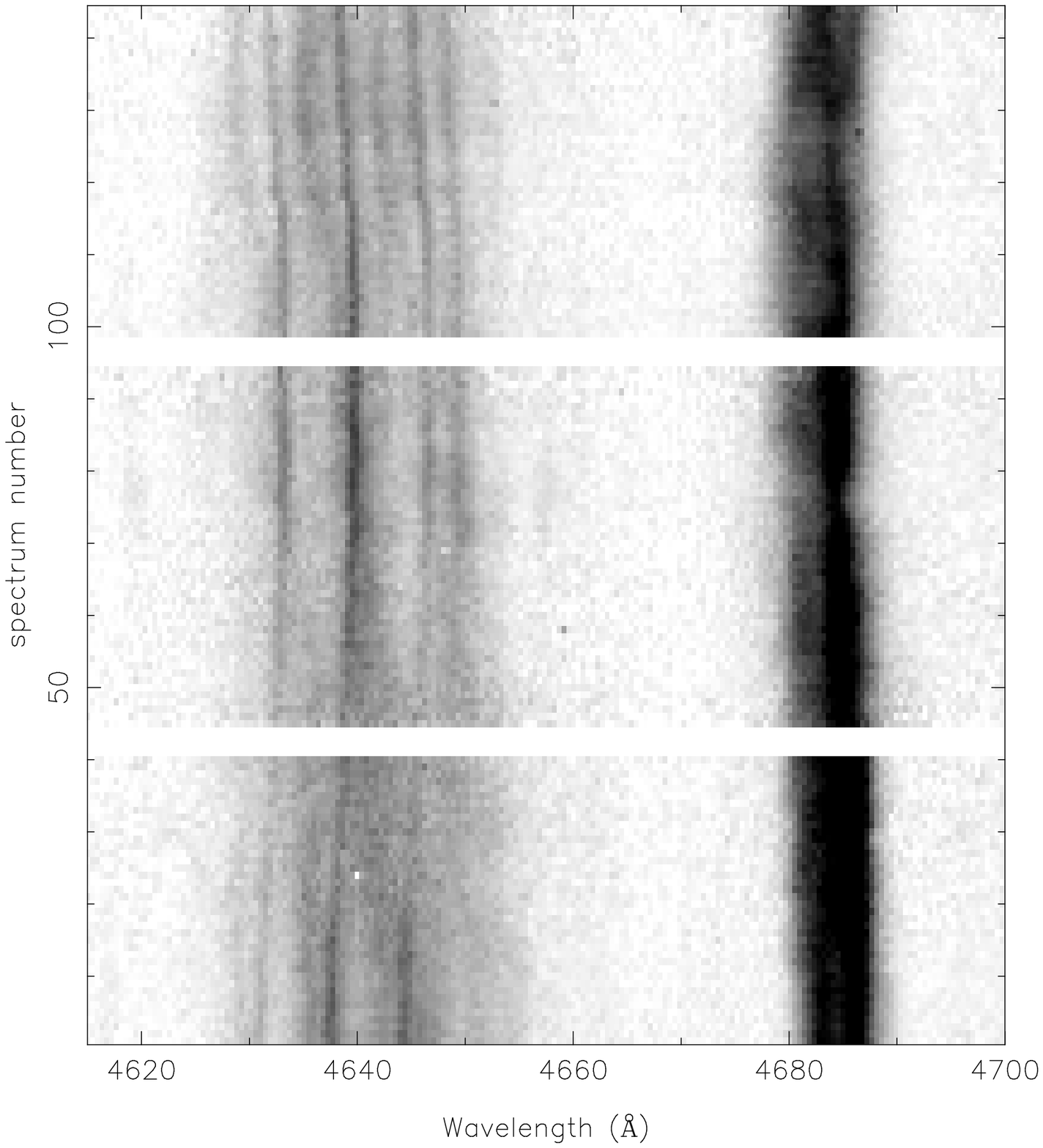}}
\put(0,0){\includegraphics{scox1_fig3.ps}}
\noindent
\end{picture}
\end{center}
\sidebyside
{\caption{Trailed spectra of the narrow CIII+NIII Bowen emission lines and HeII 
$\lambda$4686 in Sco X-1. After Steeghs \& Casares (2002).}}
{\caption{Radial velocity curve of the sharp CIII+NIII Bowen lines (top) and the
wings of HeII $\lambda$4686 (bottom) in Sco X-1.}}
\label{fig3}
\end{figure}

The radial velocity curve of the donor star can be extracted through a
combined multigaussian fit to the three CIII/NIII lines (top panel of
Fig. 4) or using more sophisticated Doppler tomography techniques (see
e.g.  ~\cite{casa03}). Furthermore, the velocities are in antiphase
with the wings of the HeII $\lambda$4686 emission, which approximately 
trace the motion of the compact star (bottom panel of Fig. 4). This work
represents the first detection of the companion star in Sco X-1 and
opens a new window for extracting dynamical information and deriving
mass functions in the population of $\sim$20 LMXBs with established
optical counterparts.

Follow-up campaigns, using the AAT, NTT and VLT telescopes, have
enabled us to extend this analysis to other fainter LMXBs leading to
the detection of the secondary stars in 2A1822-371, MXB1636-536,
MXB1735-444 and XTE J1814-338 and the determination of their orbital
velocities, which lie in the range 200-300 km s$^{-1}$ (see
~\cite{casa03}, \cite{casa04b}).  In addition, the application of this
technique to the BH candidate GX339-4 during its 2002 outburst
provided the first determination of its mass function and hence
dynamical proof that it is a BH (Hynes et al. 2003).

\begin{figure}[ht]
\begin{center}
\begin{picture}(250,130)(50,30)
\put(0,0){\includegraphics{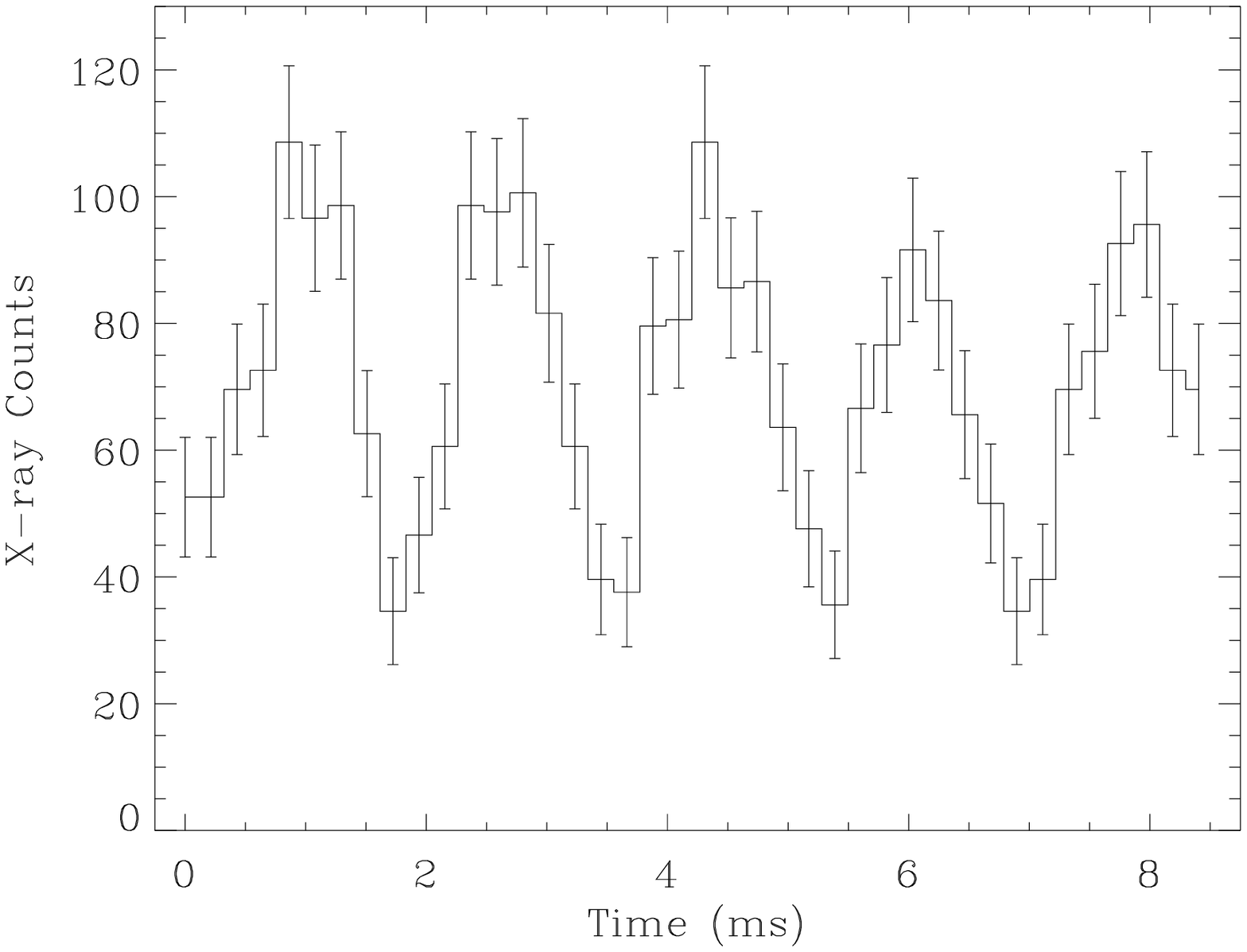}}
\put(0,0){\includegraphics{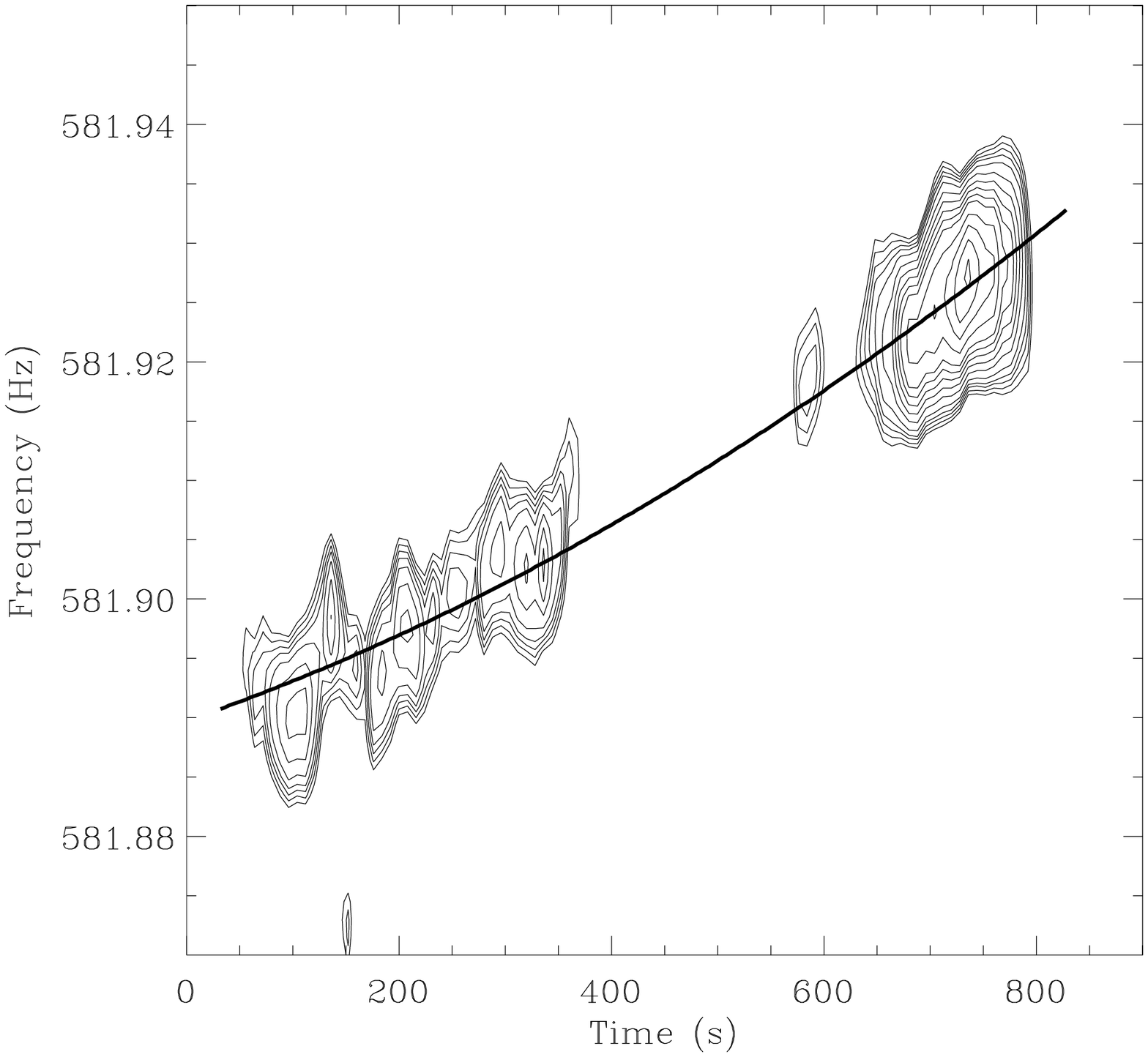}}
\noindent
\end{picture}
\end{center}
\sidebyside
{\caption{Millisecond pulses detected during the rise of an X-ray burst in
MXB1636-536 (after Strohmayer et al. 1998).}}
{\caption{Time evolution of the 580 Hz oscillation in MXB1636-536 during a 
superburst episode. The solid curve shows the best orbital model fit which
constrains the NS velocity amplitude to 90--175 km
s$^{-1}$ .}}
\label{fig2}
\end{figure}

LMXBs are considered to be the progenitors of BMPs because they provide a 
mechanism for spinning up NS to millisecond periods, through the sustained 
accretion of matter with high angular momentum during their long active lives. 
Despite intensive efforts over 2 decades, the detection of millisecond pulsations 
in LMXBs proved elusive but the advent of XTE changed things dramatically 
with the discovery of: 
(i) persistent pulses in 5 transient LMXBs with spin periods in the range 
185-435 Hz 
(ii) nearly coherent oscillations during X-ray bursts in 13 LMXBs. Figure 5 
presents an example of a train of oscillations, with a 
frequency of 580 Hz, detected during an X-ray burst in  
MXB1636-536. In two LMXBs, SAX J1808-3658 (\cite{chak03}) and XTE J1814-338 
(\cite{stro03}), burst oscillations were detected in addition 
to persistent pulses and  they showed identical frequencies. This confirmed 
that burst oscillations are indeed modulated with the spin of the NS.

Moreover, burst oscillations were detected in an 800 s interval during a  
superburst in MXB1636-536. The oscillations showed a clear frequency drift 
which was attributed to the orbital Doppler shift of the NS (see fig. 6). A 
circular orbit model was fitted to the data and used to constrain the 
projected NS velocity to between 90 and 175 km s$^{-1}$ (\cite{stro02}). 
This is a remarkable result which shows that burst oscillations can be used to 
trace NS orbits and, in combination with information provided by the Bowen 
fluorescent lines, turn persistent LMXBs into double-lined spectroscopic
binaries. 

\section{Conclusions}

In the past 15 years the field of X-ray binaries has experienced 
significant progress with the discovery of 16 new BHs and 6 millisecond 
pulsars in LMXBs. 
Reliable mass determinations have been provided for 15 BHs and 33 NS which 
are starting to reveal the mass spectrum of compact remnants. 
Two X-ray binaries, V404 Cyg and GRS 1915+105, contain BHs too massive to be  
explained by current theoretical modelling. However, better statistics are 
required in order to use the observed distribution to constrain fundamental 
parameters of X-ray binary evolution and supernova models, such as the 
mass-loss rate in the W-R phase, the explosion energy dependence on 
progenitor mass, the amount of fall back or details of the Commom Envelope 
phase.    
The discovery of fluorescence emission from the companion star, together with 
X-ray burst oscillations has opened the door to derive NS masses in persistent 
X-ray binaries. This new technique, which will benefit from new 
instrumentation on large telescopes (e.g. OSIRIS on GTC), will likely provide 
further evidence for the existence of massive NS in LMXBs.

\paragraph{Note added}
Two new NS spin periods were discovered when finishing writing this 
contribution: burst oscillations at 45 Hz in the persistent LMXB EXO 
0748-676 (\cite{villa04}) and persistent pulsations at 598.9 Hz in the 
new transient LMXB IGR J00291+5934 (\cite{mark04}), discovered by INTEGRAL 
in December 2004.

\begin{acknowledgments}
The would like to acknowledge helpful comments from my colleagues Phil Charles, 
Danny Steeghs and Tariq Shahbaz. I'm also grateful for support from the 
Spanish MCYT grant AYA2002-0036 and the programme Ramon y Cajal.
\end{acknowledgments}

\end{document}